\documentclass[reprint, prb]{revtex4-2}
\usepackage[pdftex, colorlinks, allcolors=blue]{hyperref}
\usepackage{amsthm}
\usepackage{mathtools}
\usepackage{physics}
\usepackage{graphicx}
\usepackage[T1]{fontenc}
\usepackage{siunitx}
\usepackage{comment}
\usepackage{cases}
\usepackage{multirow, makecell}

\begin{document}

\title{\texorpdfstring{Dynamic stabilization of perovskites at elevated temperatures: A comparison between cubic BaFeO\textsubscript{3} and vacancy-ordered monoclinic BaFeO\textsubscript{2.67}}{Dynamic stabilization of perovskites at elevated temperatures: A comparison between cubic BaFeO3 and vacancy-ordered monoclinic BaFeO2.67}}
\author{Yongliang Ou}
    \email{yongliang.ou@imw.uni-stuttgart.de}
\author{Yuji Ikeda}
\author{Oliver Clemens}
\author{Blazej Grabowski}
\affiliation{Institute for Materials Science, University of Stuttgart, 70569 Stuttgart, Germany}

\date{\today}


\begin{abstract}
The impact of ordered vacancies on the dynamic stability of perovskites is investigated under the \textit{ab initio} framework with a focus on cubic BaFeO$_{3}$ ($Pm\bar{3}m$) and vacancy-ordered monoclinic BaFeO$_{2.67}$ ($P2_{1}/m$). The harmonic approximation shows that both structures are dynamically unstable at \SI{0}{\kelvin}. For the monoclinic structure, the instability is related to rotational distortions of the Fe coordination tetrahedra near the ordered vacancies. \textit{Ab initio} molecular dynamics simulations in combination with the introduced structural descriptor demonstrate that both structures are stabilized above \SI{130}{\kelvin}. Our results suggest that the ordered vacancies do not significantly alter the critical temperature at which Ba--Fe--O perovskites are dynamically stabilized. Further, strong anharmonicity for the vacancy-ordered structure above its critical temperature is revealed by a significant asymmetry of the trajectories of O anions near the ordered vacancies. 
\end{abstract}

\maketitle

\section{Introduction} \label{sec:intro}

Vacancy-ordered perovskites attract increasing attention in fields such as optoelectronics~\cite{Shao2019}, photovoltaics~\cite{Ju2018}, and electrochemistry~\cite{Sengodan2015, IqbalWaidha2021} due to their tunable electronic, magnetic, and catalytic properties. The versatility of these materials is related to a stability competition among various structural arrangements. The Ba--Fe--O system offers a particularly rich class of perovskite-type structures with various vacancy-orderings, e.g., hexagonal BaFeO$_{2.65}$ ($P6_{3}/mmc$)~\cite{Gomez2001}, monoclinic BaFeO$_{2.5}$ ($P2_{1}/c$)~\cite{Clemens2014} or BaFeO$_{2.67}$ ($P2_{1}/m$)~\cite{Wollstadt2021}.

Despite considerable experimental efforts, the actual vacancy ordering and, thus, the exact arrangement of the atoms in the non-stoichiometric Ba--Fe--O phases is often not resolved~\cite{Clemens2014} on account of the structural complexity introduced by the vacancies. Experiments face practical challenges, e.g., in achieving high phase purity or in coping with multi-stage phase transitions that hinder the observation of single phases in extended temperature intervals~\cite{Wollstadt2021}. From a more fundamental perspective, it is the intricate interplay of vacancy ordering and the dynamic stability that poses a decisive challenge. 

Traditionally, the Goldschmidt tolerance factor~\cite{Goldschmidt1926} and its extended forms~\cite{Bartel2019, Sato2016, Kieslich2015}, most of which take the chemical formula and the ionic radii as the basic input data, have been used to predict the dynamic stability of perovskites. These empirical rules are, however, oversimplified and hence insufficient to reveal the details of the dynamic stability. For example, they do not capture structural deformations and, likewise, do not predict the dynamic stabilization at elevated temperatures. In order to properly take such features into account, \textit{ab initio} simulations, specifically in the form of density-functional theory (DFT), are required.

In \textit{ab initio} simulations, the harmonic approximation is commonly used to determine the dynamic (in)stability at \SI{0}{\kelvin}~\cite{Togo2015}. Indeed, for BaFeO$_3$ (i.e., the ``perfect'' cubic perovskite without ordered vacancies), a few studies have been reported to date, as summarized in Table~\ref{tab:pre}, specifically employing the DFT$+U$ approach~\cite{Dudarev1998}. DFT+$U$ is computationally much more affordable than DFT supplemented with advanced hybrid functionals, but an additional input parameter, i.e., $U_{\textrm{eff}}$, is needed to correct for the over-delocalization error of the standard exchange--correlation functionals of DFT. The specific $U_{\textrm{eff}}$ value is not trivial to determine and, as Table~\ref{tab:pre} reveals, a small change can lead to qualitatively different results. For example, cubic BaFeO$_{3}$ was predicted to be dynamically stable with $U_{\textrm{eff}} \approx \SI{5}{\electronvolt}$~\cite{Cherair2017, Zhang2018a}, while Jahn--Teller distortions were observed with $U_{\textrm{eff}} = \SI{4}{\electronvolt}$~\cite{Cherair2018, Hoedl2021}. These differences highlight that the $U_{\textrm{eff}}$ value needs to be chosen with care, ideally by calibration with respect to a higher level method. For vacancy-ordered Ba--Fe--O perovskites (i.e., compositions with lower O content than BaFeO$_3$), to our knowledge, \textit{ab initio} investigations of the dynamic (in)stability do not exist.

\begin{table}[tbp]
	\caption{Theoretical studies of the \SI{0}{\kelvin} dynamic stability of Ba--Fe--O perovskites. Cubic BaFeO$_{3}$ is ferromagnetic (FM) while monoclinic BaFeO$_{2.67}$ is G-type (in the Wollan--Koehler notation~\cite{Wollan1955}) anti-ferromagnetic (AFM). Discussions on the phonon dispersions are given in the Appendix. }
	\begin{ruledtabular}
	\begin{tabular}{cccccc}
		Year & Phase & Method & $U_{\textrm{eff}}$ (eV) & Stability \\
		\hline
		2017~\cite{Cherair2017} & cub. BaFeO$_{3}$ & DFT+\textit{U} & 5 & Stable \\
		2018~\cite{Cherair2018}  &  cub. BaFeO$_{3}$ & DFT+\textit{U} & 4 & Unstable \\
		2018~\cite{Zhang2018a} & cub. BaFeO$_{3}$ & DFT+\textit{U} & 5.2 & Stable \\
		2021~\cite{Hoedl2021} & cub. BaFeO$_{3}$ & DFT+\textit{U} & 4 & Unstable \\
		this study & cub. BaFeO$_{3}$ & HSE06 & N/A & Unstable \\
		this study & cub. BaFeO$_{3}$ & DFT+\textit{U} & 3 & Unstable \\
		this study & mon. BaFeO$_{2.67}$ & DFT+\textit{U} & 3 & Unstable \\
	\end{tabular}
	\end{ruledtabular}
	\label{tab:pre}
\end{table}

Dynamic stabilization at finite temperatures can be investigated via \textit{ab initio} molecular dynamics (AIMD) simulations. One type of approaches utilizes effective force constants extracted from AIMD. For example, the temperature-dependent effective potential (TDEP) method~\cite{Hellman2011} fits effective harmonic force constants to the forces generated by AIMD. With the self-consistent phonon (SCPH) method, Tadano \textit{et al.}~\cite{Tadano2015} showed how to obtain renormalized phonon frequencies from anharmonic force constants extracted from AIMD. In general, sufficiently long AIMD runs are required for these methods to obtain reliable results, especially for materials with significant anharmonicity. Another strategy is to analyze the lattice distortions of the structures directly with AIMD. The dynamic stabilization can be then visualized explicitly as a function of temperature, and the anharmonic contribution fully considered. Such an approach was applied to cubic perovskites~\cite{Klarbring2018, Sun2014}, though not yet to vacancy-ordered perovskites. 

The aim of the present study is to compare the dynamic stability of cubic BaFeO$_{3}$ ($Pm\bar{3}m$)~\cite{Hayashi2011} (no ordered vacancies) with vacancy-ordered monoclinic BaFeO$_{2.67}$ ($P2_{1}/m$)~\cite{Wollstadt2021} at \SI{0}{\kelvin} and at finite temperatures with AIMD and, thereby, to reveal the impact of ordered vacancies on the dynamic stability of perovskites. We first demonstrate that $U_\mathrm{eff} = \SI{3}{eV}$ should be chosen, since it well reproduces the dynamic instability predicted by the more accurate HSE06 functional~\cite{Krukau2006, Wahl2008}. Based on the optimized $U_\mathrm{eff}$, we investigate the link between the ordered vacancies and the imaginary harmonic phonon modes at \SI{0}{\kelvin}. We analyze trajectories of the atoms near the ordered vacancies at finite temperatures and investigate the influence of the ordered vacancies on the stabilization mechanism. To this end, we introduce a structural descriptor that captures the temperature-driven transformation for both, the vacancy-free cubic and vacancy-ordered monoclinic structures.
\section{Methodology} \label{sec:metho}

\subsection{Cubic and monoclinic structures}\label{sec:str}

\begin{figure*}[tbp]
    \centering
    \includegraphics[width=17.8cm]{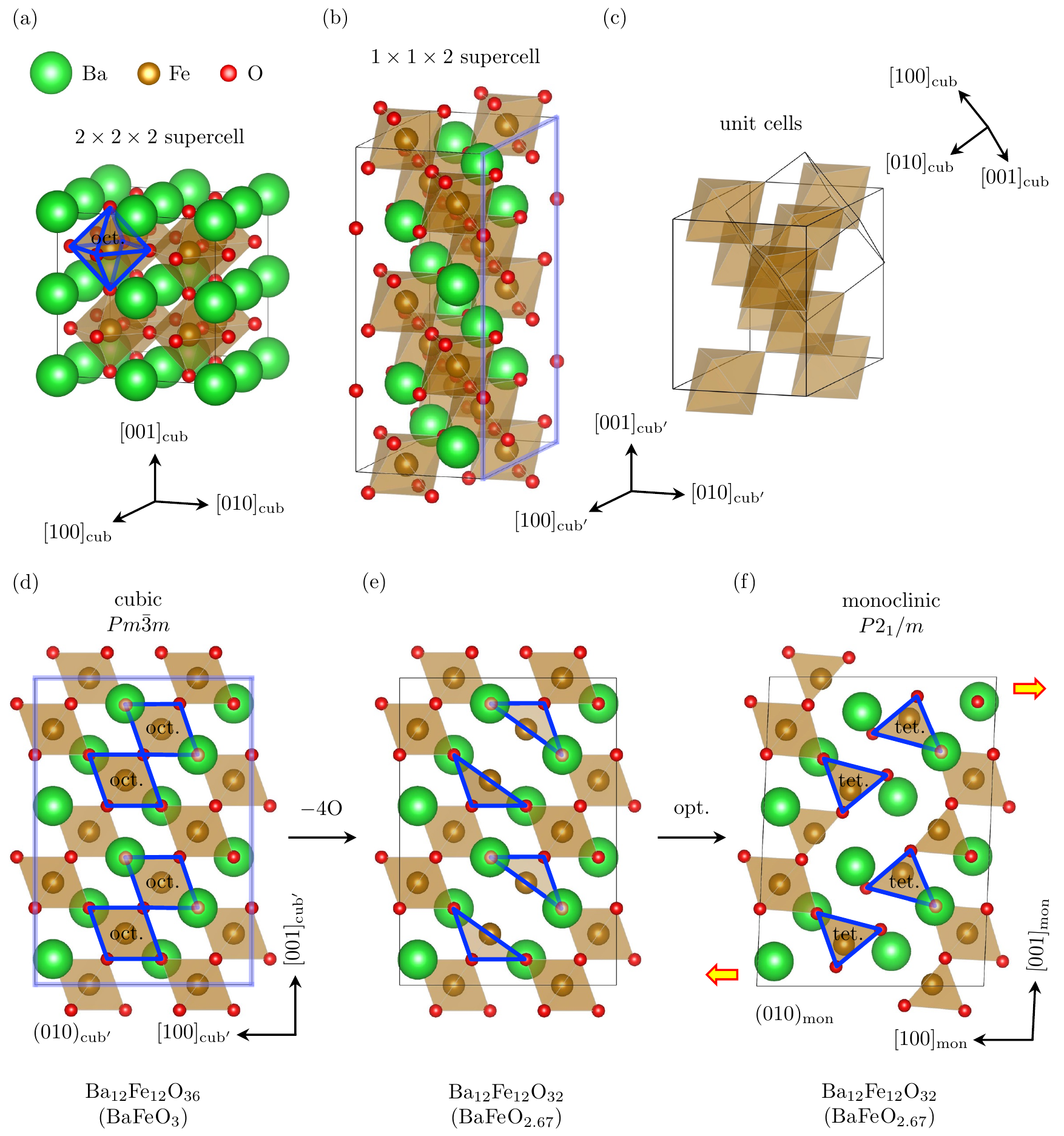}
    \caption{Derivation of the vacancy-ordered monoclinic structure from the cubic structure. (a) Simulation cell (\num{2 x 2 x 2} supercell) of the cubic structure. (b) Simulation cell (\num{1 x 1 x 2} supercell) after coordinate transformation. (c) Orientation comparison of the original and the transformed unit cells. (d) $(010)_{\textrm{cub}'}$ plane projection of the supercell (b) with the four highlighted octahedra. (e) Vacancy creation on the O sites. (f) Simulation cell (\num{1 x 1 x 2} supercell) of the monoclinic structure, with indication of the shearing during the structural optimization process. The numbers of atoms (formula unit) in the cells are shown below for (d)--(f). Crystal structures are visualized by \textsc{vesta}~\cite{Momma2011}.}
    \label{fig:trans}
\end{figure*}

Figure~\ref{fig:trans}(a) shows the utilized simulation cell of the ideal cubic BaFeO$_{3}$ structure. While the unit cell contains 5 atoms, in the present study, a \num{2 x 2 x 2} supercell (40 atoms), which contains eight symmetrically equivalent corner-shared regular Fe coordination octahedra, was considered to capture the low temperature distortion of the structure. Each octahedron consists of one central Fe atom and six surrounding O atoms. Ba atoms are located between the octahedra.

To derive the vacancy-ordered structure, a coordinate transformation~\cite{Clemens2015, Wollstadt2021, IqbalWaidha2021} is performed on the unit cell of the cubic structure. The lattice vectors of the transformed unit cell $\boldsymbol{a}_{\textrm{cub}'}$, $\boldsymbol{b}_{\textrm{cub}'}$ and $\boldsymbol{c}_{\textrm{cub}'}$ are related to that of the original unit cell $\boldsymbol{a}_{\textrm{cub}}$, $\boldsymbol{b}_{\textrm{cub}}$ and $\boldsymbol{c}_{\textrm{cub}}$ according to
\begin{equation}
    \begin{pmatrix}
    \boldsymbol{a}_{\textrm{cub}'} \\
    \boldsymbol{b}_{\textrm{cub}'} \\
    \boldsymbol{c}_{\textrm{cub}'}
    \end{pmatrix} = 
    \begin{pmatrix}
    1 & -1 & 0 \\
    1 & 1 & 1 \\
    -1 & -1 & 2
    \end{pmatrix}
    \begin{pmatrix}
    \boldsymbol{a}_{\textrm{cub}} \\
    \boldsymbol{b}_{\textrm{cub}} \\
    \boldsymbol{c}_{\textrm{cub}}
    \end{pmatrix}.
\end{equation}
While this transformation leads to a six-times larger unit cell, to simulate the dynamic stability of the vacancy-ordered structure, as well as to realize the G-type AFM ordering (see Sec.~\ref{sec:comp} for details), it was further expanded by \num{1 x 1 x 2}. The thus obtained simulation cell contains 12 corner-shared regular Fe coordination polyhedra, as shown in Fig.~\ref{fig:trans}(b). The orientation relationship between the two unit cells is shown in Fig.~\ref{fig:trans}(c). 

The $(101)_{\textrm{cub}'}$ plane projection of the transformed cubic structure with four highlighted octahedra is shown in Fig.~\ref{fig:trans}(d). Ordered vacancies are created by removing the four O atoms belonging to these four octahedra, as shown in Fig.~\ref{fig:trans}(e). The composition of the supercell changes from Ba$_{12}$Fe$_{12}$O$_{36}$ to Ba$_{12}$Fe$_{12}$O$_{32}$, i.e., the formula unit (f.u.) changes from BaFeO$_{3}$ to BaFeO$_{2.67}$. After the formation of the vacancies, four of the twelve previously octahedrally coordinated Fe cations become tetrahedrally coordinated. The symmetry of the structure is lowered, and the space group is changed to monoclinic $P2_{1}/m$. As a final step, the vacancy-ordered structure is optimized in an \textit{ab initio} manner, during which relaxation of the four tetrahedra and a shear of the supercell in the $[100]_{\textrm{mon}}$ direction take place, as shown in Fig.~\ref{fig:trans}(f).

\subsection{\texorpdfstring{Structural descriptor $\Delta$}{Structural descriptor Delta}}\label{sec:strdes}

\begin{figure}[tbp]
    \centering
    \includegraphics[width=8.6cm]{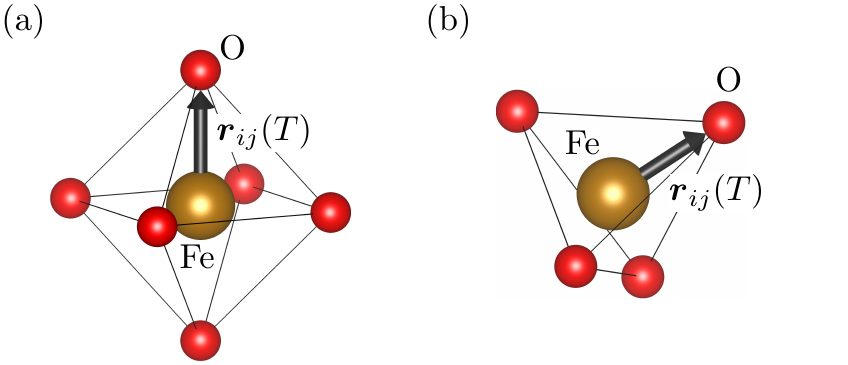}
    \caption{Illustration of relative position vectors in (a) the octahedron and (b) the tetrahedron cases. }
    \label{fig:fe_o}
\end{figure}

To quantify the displacive phase transformation at the atomistic level, we introduce a structural descriptor $\Delta$ along similar lines as done previously for materials showing the bcc--$\omega$ phase transformation~\cite{Korbmacher2019, Ikeda2021, Gubaev2021}. The key requirements are as follows:
\begin{itemize}
\item The structural descriptor $\Delta(T)$ is a temperature dependent, scalar value that condenses the relevant information on the displacive phase transformation from AIMD trajectories.
\item Thermal vibrations (i.e., random displacements) are filtered out to a good degree to increase the contrast in $\Delta$ between the low-temperature, low-symmetry phase and the high-temperature, high-symmetry phase.
\item Displacements of atoms unrelated to symmetry breaking are filtered out effectively as well as similarly for structures restricted by different space groups.
\item The vacancy-free cubic and the vacancy-ordered monoclinic structure are treated on an equal footing, i.e., O anions in octahedral environments are similarly considered as O anions in tetrahedral environments.
\end{itemize}
To fulfill these requirements we define $\Delta$ as follows. For a polyhedron composed of a central Fe cation $i$ and surrounding O anions labelled with $j$, the temperature dependent relative position vectors $\boldsymbol{r}_{ij}(T)$ are defined as
\begin{equation}
    \boldsymbol{r}_{ij}\left(T\right) = \left\langle\boldsymbol{R}_{j}\right\rangle_{T} - \left\langle\boldsymbol{R}_{i}\right\rangle_{T},
    \label{eq:posvec}
\end{equation}
where $\boldsymbol{R}_i$ is the position vector of the $i$th cation and $\boldsymbol{R}_j$ of the $j$th anion. The time-averaged position of an ion at a given temperature $T$ is represented by $\left<\cdots\right>_{T}$. Figures~\ref{fig:fe_o}(a) and \ref{fig:fe_o}(b) give examples of the relative position vectors $\boldsymbol{r}_{ij}$ in the octahedron and the tetrahedron cases, respectively. Based on the position vectors $\boldsymbol{r}_{ij}$, we define a structure-dependent projection as 
\begin{numcases}{p_{ij} = }
\boldsymbol{r}_{ij} \cdot \hat{\boldsymbol{r}}_{ij} & cubic, \label{eq:cub}
\\
\boldsymbol{r}_{ij} \cdot \hat{\boldsymbol{u}}_{[010]_{\textrm{mon}}} & monoclinic, \label{eq:mon}
\end{numcases}
which gives a parameter $p_{ij}$. The relative position vector $\boldsymbol{r}_{ij}$ is projected onto the unit vectors in the same direction of $\boldsymbol{r}_{ij}$ ($\hat{\boldsymbol{r}}_{ij}$) for the cubic structure and in the symmetry broken direction ($\hat{\boldsymbol{u}}_{[010]_{\textrm{mon}}}$) for the monoclinic structure.

To emphasize the distortion, the parameter of the ideal structure $p^{\textrm{ideal}}$ is subtracted from the simulated parameters $p_{ij}$ at different temperatures. The structural descriptor is obtained as the mean value of the referenced $p_{ij}$ parameters of the cation-anion pairs inside the polyhedra, 
\begin{equation}
    \Delta = \frac{1}{I\times J} \sum_{i = 1}^{I} \sum_{j = 1}^{J} \left|p_{ij} - p^{\textrm{ideal}}\right|,
    \label{eq:delta}
\end{equation}
where $I$ is the number of cations and $J$ is the number of anions surrounding each cation.

The ideal cubic structure belongs to the space group $Pm\bar{3}m$ (No.~221), and all O atoms are at the Wyckoff site $3c$ for which all three fractional coordinates are fixed. This means that the space group of the cubic structure restricts all O atoms to a single spatial point in the simulation cell. The elongation or contraction of a cation--anion bonding in all directions belongs to the distortion of the cubic structure. Therefore, the relative position vector $\boldsymbol{r}_{ij} (T)$ is projected to itself, i.e., the module of $\boldsymbol{r}_{ij}$ is calculated [Eq.~\eqref{eq:cub}]. All Fe--O bondings in each of the octahedra in the simulation cell are taken into account for Eq.~\eqref{eq:delta}.

For the monoclinic structure, not all atoms are spatially restricted by the symmetry operations included in the space group, i.e., the structure can still belong to the same space group $P2_{1}/m$ as long as the displacements of atoms do not break the symmetry. As will be discussed in Sec.~\ref{sec:mon}, the dynamic instability of the monoclinic structure is mainly related to the distortion of the four tetrahedra, in which only the four O atoms moving in the $[010]_{\textrm{mon}}$ direction contribute to breaking of the symmetry. To investigate the dynamic instability of $P2_{1}/m$ BaFeO$_{2.67}$, we therefore consider the Fe--O pairs [Fig.~\ref{fig:fe_o}(b)] involving the four O atoms for estimating the distortion and calculating $\Delta$. Mathematically, the four $\boldsymbol{r}_{ij} (T)$ [Eq.~\eqref{eq:posvec}] are projected to the unit vector along the $[010]_{\textrm{mon}}$ direction [Eq.~\eqref{eq:mon}], and $I=4, J=1$ in Eq.~\eqref{eq:delta}.

\subsection{Computational details}\label{sec:comp}

Spin alignment is important for Fe oxides. Experiments revealed that cubic BaFeO$_{3}$ has an A-type spiral spin structure near \SI{0}{\kelvin}~\cite{Hayashi2011}, which was subsequently investigated by \textit{ab initio} simulations~\cite{Li2012a}. Further, a magnetic transition to the FM state was observed experimentally at \SI{111}{\kelvin}~\cite{Hayashi2011}. The FM state was also shown to be energetically the most stable within calculations with collinear spin alignment~\cite{Ribeiro2013, Maznichenko2016, Rahman2016}. For the vacancy-ordered monoclinic BaFeO$_{2.67}$ structure, experimental and \textit{ab initio} results show agreement on a G-AFM ordering~\cite{Wollstadt2021}. In the present study, collinear spin alignment was applied, and the FM and the G-AFM states were considered for the cubic and the monoclinic phases, respectively.

Electronic structure calculations were carried out under the DFT framework using the projector augmented wave (PAW) method~\cite{Bloechl1994} and the generalized gradient approximation (GGA) in the Perdew--Burke--Ernzerhof (PBE) parametrization~\cite{Perdew1996} as implemented in \textsc{vasp}~\cite{Kresse1995, Kresse1996, Kresse1999}. Electrons in the atomic orbitals $5s^{2}5p^{6}6s^{2}$ (Ba), $3d^{6}4s^{2}$ (Fe) and $2s^{2}2p^{4}$ (O) were treated as valance electrons. For calculations using the DFT$+U$ method, the effective Hubbard potential ($U_{\textrm{eff}}$)~\cite{Dudarev1998} was added to electrons in the $d$-orbitals of the Fe atoms to capture the strong on-site Coulomb interaction. Additionally, DFT calculations with the HSE06 hybrid functional~\cite{Krukau2006} were performed. The first-order Methfessel--Paxton scheme~\cite{Methfessel1989} with a smearing width of \SI{0.1}{\electronvolt} was used for structural optimization and force calculations, and the tetrahedron method with Blöchl corrections~\cite{Bloechl1994a} was used for accurate energy calculations. The plane-wave cutoff was set to \SI{520}{\electronvolt}. The reciprocal space was sampled by $\Gamma$-centered \num{4 x 4 x 4} and \num{3 x 6 x 3} $\boldsymbol{k}$-point meshes for the 40-atom cubic BaFeO$_{3}$ and the 54-atom monoclinic BaFeO$_{2.67}$ simulation cells, respectively. For the Kohn--Sham self-consistent calculation, the energy was minimized until the energy difference converged to less than \SI{d-5}{\electronvolt} per simulation cell. Ionic relaxation was performed with the conjugate gradient algorithm until the maximum residual force was less than \SI{d-2}{\electronvolt\angstrom{}^{-1}}.

The \SI{0}{\kelvin} harmonic phonon dispersions were calculated using the finite displacement method implemented in \textsc{phonopy}~\cite{Togo2015}. The displacement amplitude was set to \SI{d-2}{\angstrom}, and a $\Gamma$-centered \num{14 x 14 x 14} $\boldsymbol{q}$-point mesh was used for sampling the reciprocal space. Tests show that better energy convergences do not significantly alter the results, so the current criterion (\SI{d-5}{\electronvolt} per simulation cell) was also used for the phonon calculations. The residual forces in the optimized structure were subtracted from the forces of the displaced structure for accurate calculations of the interatomic force constants. 

AIMD simulations were conducted using the Langevin thermostat and the canonical ensemble implemented in \textsc{vasp}~\cite{Kresse1995, Kresse1996, Kresse1999}. A \SI{2}{\femto\second} time step, a \SI{10}{\pico\second^{-1}} friction coefficient and the Fermi--Dirac smearing adjusted to the MD temperature were utilized. For each AIMD step, the criterion for energy convergence was set to \SI{d-3}{\electronvolt} per simulation cell, and the atomic positions were calibrated to fix the center of mass. Other parameters were the same as those chosen for the electronic structure calculations. In total 5000 AIMD steps (\SI{10}{\pico\second}) were performed, and the first 300 steps were subtracted for thermalization. 


\section{Results}\label{sec:result}

\subsection{\texorpdfstring{Cubic BaFeO\textsubscript{3}}{Cubic BaFeO3}} \label{sec:cubic}

\begin{figure*}[tbp]
    \centering
    \includegraphics[width=17.8cm]{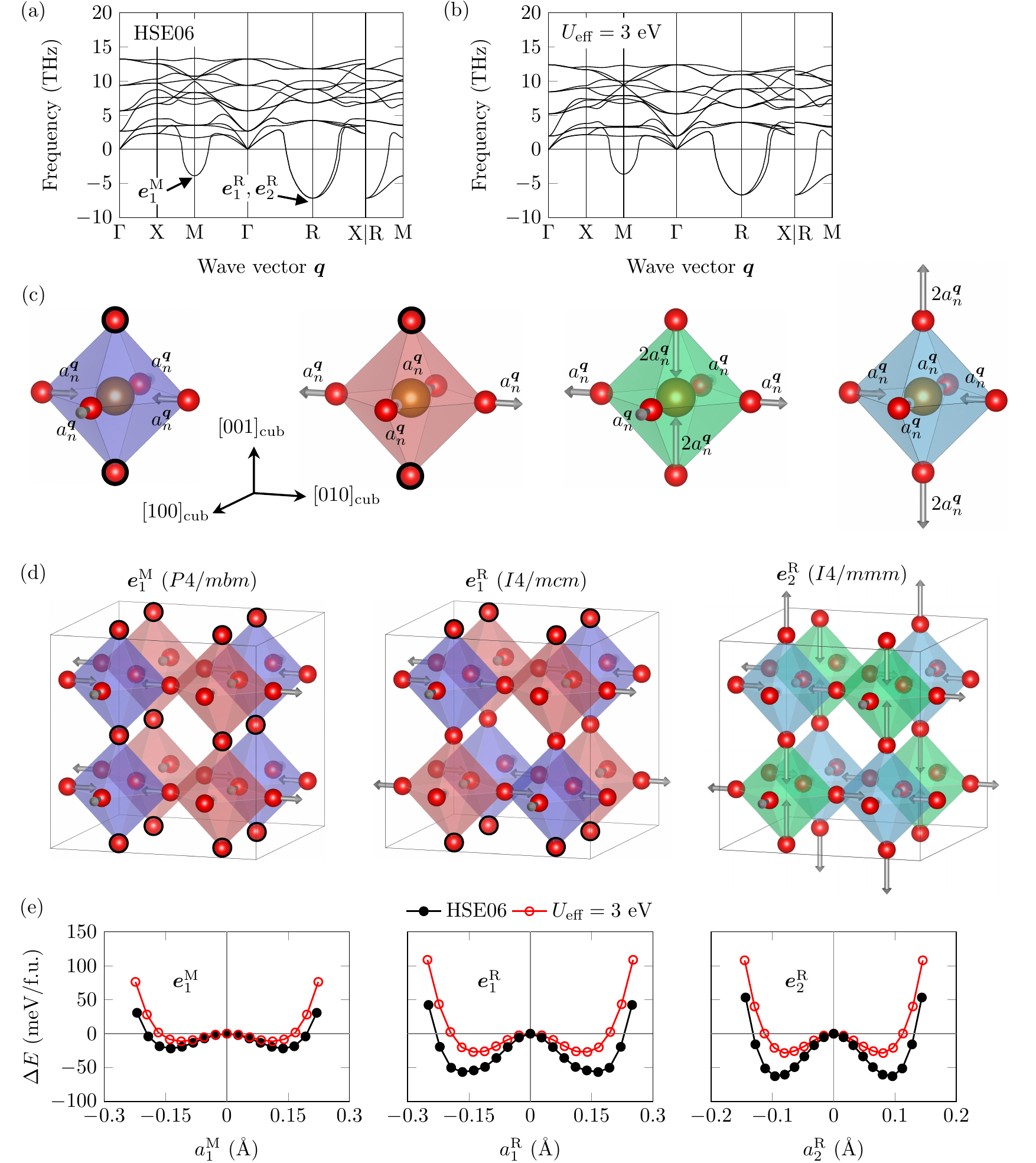}
    \caption{Dynamic instability of the cubic structure at \SI{0}{\kelvin}. Phonon dispersions calculated by (a) HSE06 and (b) DFT$+U$ ($U_{\textrm{eff}} = \SI{3}{\electronvolt}$). Commensurate $\boldsymbol{q}$-points are indicated by tick labels. Negative frequencies indicate the imaginary modes. (c) Four distortion types of an octahedron. Arrows indicate the movement of the atoms. Stationary atoms are anchored by black circles. (d) Collective motion of atoms for the three imaginary modes indicated in (a). (e) The double-well potentials of the three imaginary modes. The corresponding displacements of the O atoms are indicated in (c).}
    \label{fig:pho3}
\end{figure*}

Figure~\ref{fig:pho3}(a) shows the \SI{0}{\kelvin} phonon dispersion calculated by HSE06 for the cubic BaFeO$_{3}$ structure. As indicated by the arrows, one imaginary mode $\boldsymbol{e}^{\textrm{M}}_{1}$ at the M point [$\boldsymbol{q} = \left(1/2, 1/2, 0\right)$] and two degenerate imaginary modes $\boldsymbol{e}^{\textrm{R}}_{1}$ and $\boldsymbol{e}^{\textrm{R}}_{2}$ at the R point [$\boldsymbol{q} = \left(1/2, 1/2, 1/2\right)$] are observed, which reveals dynamic instability for the cubic structure at \SI{0}{\kelvin}. The imaginary modes correspond to collective displacements on the O atom sublattice. These displacements are related to the Jahn--Teller effect~\cite{Cherair2018, Hoedl2021} and they deform the octahedra in specific ways as illustrated in Figs.~\ref{fig:pho3}(c) and~\ref{fig:pho3}(d). There are, for example, breathing-type displacements, in which the O atoms in one plane move simultaneously inward or outward. The displacements cause a decrease of the symmetry, i.e., the space group is changed from cubic $Pm\bar{3}m$ to tetragonal $P4/mbm$, $I4/mcm$ and $I4/mmm$ for distortions along the $\boldsymbol{e}_{1}^{\textrm{M}}$, $\boldsymbol{e}_{1}^{\textrm{R}}$ and $\boldsymbol{e}_{2}^{\textrm{R}}$ imaginary modes, respectively.

The double-well potentials corresponding to the imaginary modes are shown in Fig.~\ref{fig:pho3}(e) (black for HSE06). They reveal that the minimum in energy is reached already at a sub-ångström level for any of the three imaginary modes. It can be also observed that the potential wells are rather shallow (few tenths of meV/f.u.~which translates to a few meV/atom) such that the O anions should be able to overcome the potential barrier by thermal energy already at a low temperature. As shown later, AIMD simulations do confirm this statement.

For the other simulations in this study (AIMD, vacancy-ordered structure), HSE06 can hardly be used due to its high computational cost, a fact that motivates the application of the computationally cheaper DFT+$U$ approach. In the DFT$+U$ method, dynamic stability of the Ba--Fe--O perovskites depends on the input parameter $U_\mathrm{eff}$, as discussed in Sec.~\ref{sec:intro}. By comparing calculated and experimental oxidation energies, Wang~\textit{et al.}~\cite{Wang2006} recommended $U_{\mathrm{eff}}\approx\SI{4}{\electronvolt}$, which, however, is not necessarily suitable for dynamic stability investigations. In the present study, the $U_{\textrm{eff}}$ parameter has been calibrated by the just discussed HSE06 results. A range of $U_{\textrm{eff}}$ values was tested with a focus on the phonon dispersion of the FM cubic structure (results are shown in the Appendix). The best match between the two methods is obtained with $U_{\textrm{eff}} = \SI{3}{\electronvolt}$, which is close to the value used by Wollstadt \textit{et al.}~\cite{Wollstadt2021}. A comparison between Figs~\ref{fig:pho3}(a) and~\ref{fig:pho3}(b) exemplifies the good agreement for the phonon dispersion. The qualitative dependence of the imaginary branches is similar, with quantitative differences of about \SI{1}{\THz}. The reasonable agreement between the potential energies [Fig.~\ref{fig:pho3}(e) red~vs.~black] further supports the usage of $U_{\textrm{eff}} = \SI{3}{\electronvolt}$. DFT$+U$ predicts a similar width and about half of the depth of the potential wells as compared with HSE06. Since the absolute energy difference at the lowest point of the potential well is quite small (less than \SI{50}{\milli\electronvolt\per{f.u.}}), the DFT$+U$ method with $U_{\textrm{eff}} = \SI{3}{\electronvolt}$ is considered acceptable for the dynamic stability analysis.

\begin{figure*}[tbp]
    \centering
    \includegraphics[width=17.8cm]{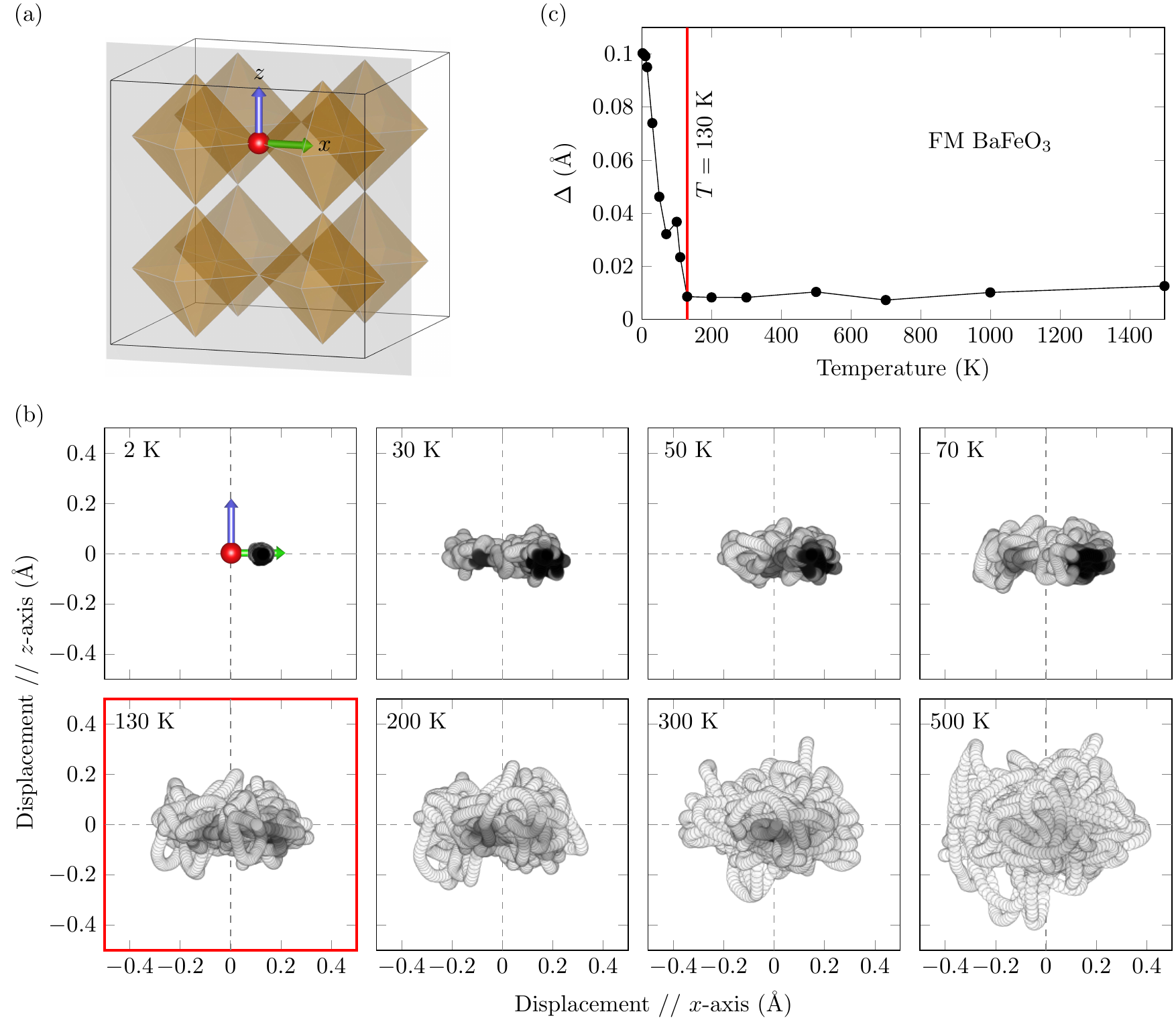}
    \caption{AIMD simulation of the FM cubic BaFeO$_{3}$. (a) Indication of one specific O atom. (b) Trajectories of the O anion at elevated temperatures. (c) The change of the structural descriptor at an elevated temperature. The DFT$+U$ method with $U_{\textrm{eff}} = \SI{3}{\electronvolt}$ was used. }
    \label{fig:cubmd}
\end{figure*}

To investigate the dynamic stability at elevated temperatures, AIMD simulations have been performed from \SI{2}{\kelvin} up to \SI{1500}{\kelvin} for the FM cubic BaFeO$_{3}$. At low temperatures, large displacements are observed for the O anions, as compared to the Ba or Fe cations, which reinforces the dynamic instability. The instability falls into the imaginary-mode regime identified from the phonon calculations, in which only O atoms show displacements [Figs.~\ref{fig:pho3}(c) and~\ref{fig:pho3}(d)]. Further relaxation (with fixed cell shape and volume) of the low-temperature distorted structure shows a decrease of the energy by \SI{-28.5}{\milli\electronvolt\per{f.u.}} as compared to the ideal cubic structure, which is the same as the energy minimum of the potential wells of the $\boldsymbol{e}_{2}^{\textrm{R}}$ mode [red lines in Fig.~\ref{fig:pho3}(e), cf.~\SI{-26.9}{\milli\electronvolt\per{f.u.}} for $\boldsymbol{e}_{1}^{\textrm{R}}$ and \SI{-11.2}{\milli\electronvolt\per{f.u.}} for $\boldsymbol{e}_{1}^{\textrm{M}}$]. Since all O atoms within the cubic structure are symmetrically equivalent, one specific O atom is used for demonstration, as depicted in Fig.~\ref{fig:cubmd}(a). The trajectories of the O anion at various temperatures are displayed in Fig.~\ref{fig:cubmd}(b). At \SI{2}{\kelvin}, the O anion is trapped at a displaced position in the $x$ direction with a distance of about \SI{0.1}{\angstrom}. The potential barrier is overcome at about \SI{30}{\kelvin}, and a relatively homogeneous distribution of the O anion is observed at \SI{130}{\kelvin} and at higher temperatures.

To quantitatively describe the transformation observed in AIMD, we utilize the structure descriptor $\Delta$ introduced in Sec.~\ref{sec:strdes}.  Figure~\ref{fig:cubmd}(c) shows the change of $\Delta$ as a function of temperature. A sharp decrease of $\Delta$ is observed just below \SI{130}{\kelvin}, while it remains almost constant at higher temperatures. The AIMD simulations thus predict that the FM cubic structure is stabilized at about \SI{130}{\kelvin} due to vibrational entropy. The local magnetic moment of Fe for the ideal cubic structure at \SI{0}{\kelvin} is $\SI{3.7}{\mu_{\textrm{B}}}$ per Fe ion. For all simulated temperatures, the FM spin state remains unchanged. 




\subsection{\texorpdfstring{Monoclinic BaFeO\textsubscript{2.67}}{Monoclinic BaFeO2.67}} \label{sec:mon}

\begin{figure*}[tbp]
    \centering
    \includegraphics[width=17.8cm]{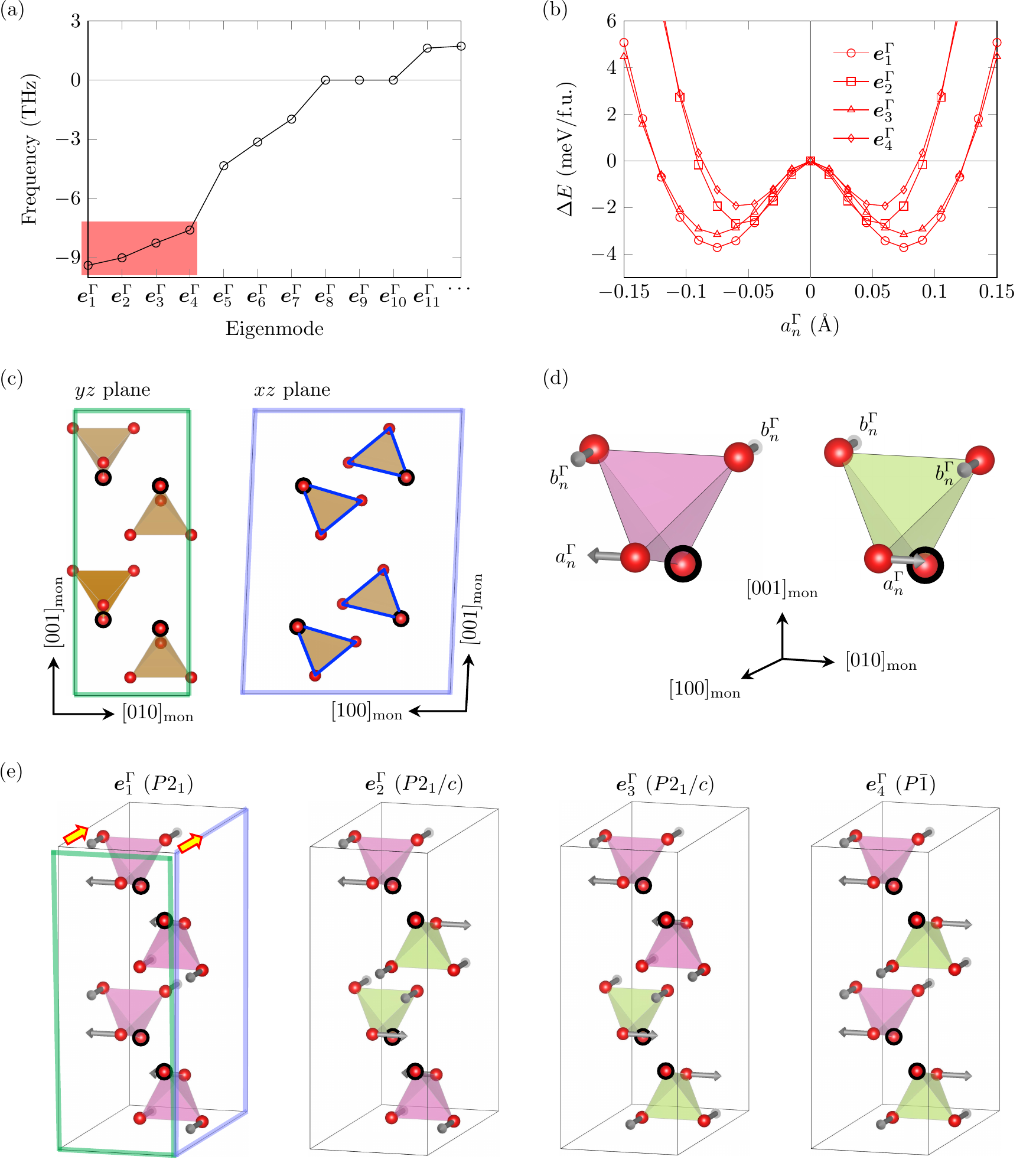}
    \caption{Dynamic instability of the monoclinic structure at \SI{0}{\kelvin}. (a) Frequencies of the first few eigenmodes at the commensurate $\Gamma$ point for the G-AFM monoclinic structure. The first four imaginary modes (highlighted in red) are of relevance for the \SI{0}{\kelvin} dynamical instability. (b) The double-well potentials of these modes. The corresponding displacements of the O atoms are indicated in (d). (c) Projections of the optimized monoclinic structure [Fig.~\ref{fig:trans}(f)] onto the $yz$ and $xz$ planes. Only the four highlighted tetrahedra and the O atoms forming them are displayed. (d) The two distortion types of a tetrahedron. Arrows indicate the movement of the atoms. Stationary atoms are anchored by black circles. (e) Collective motion of atoms for the four highlighted imaginary modes. Shearing of the simulation cell during the structural optimization process [Fig.~\ref{fig:trans}(f)] is indicated.}
    \label{fig:pho2.67}
\end{figure*}

The phonon dispersion of the G-AFM monoclinic BaFeO$_{2.67}$ structure with the symmetry of $P2_1/m$ [Fig.~\ref{fig:trans}(f)], obtained from the construction process described in Sec.~\ref{sec:str}, has been calculated at \SI{0}{\kelvin} with the DFT$+U$ method ($U_{\textrm{eff}} = \SI{3}{\electronvolt}$). Figure~\ref{fig:pho2.67}(a) shows frequencies of the first few eigenmodes at the $\Gamma$ point (the only commensurate $\boldsymbol{q}$-point) in ascending order. Seven imaginary modes are found, which clearly reveals dynamic instability of the $P2_1/m$ monoclinic structure at \SI{0}{\kelvin}. Note that the relaxation applied during the construction process cannot ``remove'' the dynamic instability observed in Fig.~\ref{fig:pho2.67}(a) due to symmetry constraints. The monoclinic structure after the relaxation corresponds to a saddle point on the potential energy surface, and a vibrational mode analysis as performed here is necessary to detect the dynamic instability.

The analysis of the phonon eigenvectors shows that the first four imaginary modes [red shaded in Fig.~\ref{fig:pho2.67}(a)] mainly correspond to displacements of several specific O anions. The heavier Ba and Fe cations are, instead, involved in the other (three) imaginary modes at less negative frequencies (about~\SI{-3}{\THz}). Since the lighter O atoms move faster than the Ba and Fe atoms, it seems reasonable to assume that the monoclinic structure is destabilized along a combination of the lowest four imaginary modes as the temperature is lowered. This assumption is indeed confirmed by the AIMD simulations (discussed in more detail below), which reveal that over $90\%$ of the low temperature distortion is contributed by these four lowest imaginary modes. We can therefore also deduce that displacements along the lowest four imaginary modes stabilize the other three imaginary modes involving Ba and Fe atom displacements through phonon--phonon interactions.

The four lowest imaginary modes are visualized in real space in Fig.~\ref{fig:pho2.67}(c)--(e). Only the relevant parts of the monoclinic structure are displayed, specifically the four tetrahedra with their corresponding O atoms. These are the essential components describing the low-temperature distortion. The four lowest imaginary modes can be built up from two types of symmetrically related deformations of the tetrahedra [Fig.~\ref{fig:pho2.67}(d)]. We recall that the tetrahedra are a result of the introduction of the ordered vacancies and the subsequent relaxation [Figs.~\ref{fig:trans}(d)--(f)].

The ideal monoclinic BaFeO$_{2.67}$ belongs to the space group $P2_{1}/m$ (No.~11). During the transformation of a tetrahedron, the O atom at the Wyckoff site $2e$ is displaced by $a_{n}^{\Gamma}$ along the $[010]_{\textrm{mon}}$ direction, which breaks the reflection symmetry with respect to the $(010)_\mathrm{mon}$ plane and thus lowers the symmetry from $P2_1/m$ to, e.g., $P2_1$. The two O atoms of the tetrahedron located at the Wyckoff site $4f$ move along the $[100]_{\textrm{mon}}$ direction, each by the same distance $b_{n}^{\Gamma}$ but in opposite direction to each other. Their positions are not restricted by the space group. No quantitative relationship between $a_{n}^{\Gamma}$ and $b_{n}^{\Gamma}$ can be derived based on the analysis of the imaginary phonon modes. 

Figure~\ref{fig:pho2.67}(b) shows the potential energy along the four lowest imaginary modes. The double-well potentials affirm the dynamic instability of the ideal monoclinic structure at \SI{0}{\kelvin}. These potential wells are shallower than the ones of the cubic structure [Fig.~\ref{fig:pho3}(e)], which, however, does not immediately mean that the monoclinic structure is easier to be stabilized at elevated temperatures. The true local minimum of the distorted monoclinic structure is not captured by displacements along single modes. Further relaxation of the low temperature distorted structure (obtained from AIMD, which includes a combination of the imaginary modes) gives a lower energy minimum of \SI{-16.1}{\milli\electronvolt\per{f.u.}} at larger O atom displacements (about $\SI{0.2}{\angstrom}$). Additionally, the energy scales of the cubic and the monoclinic structures are not directly comparable because of the different formula units. 

\begin{figure*}[tbp]
    \centering
    \includegraphics[width=17.8cm]{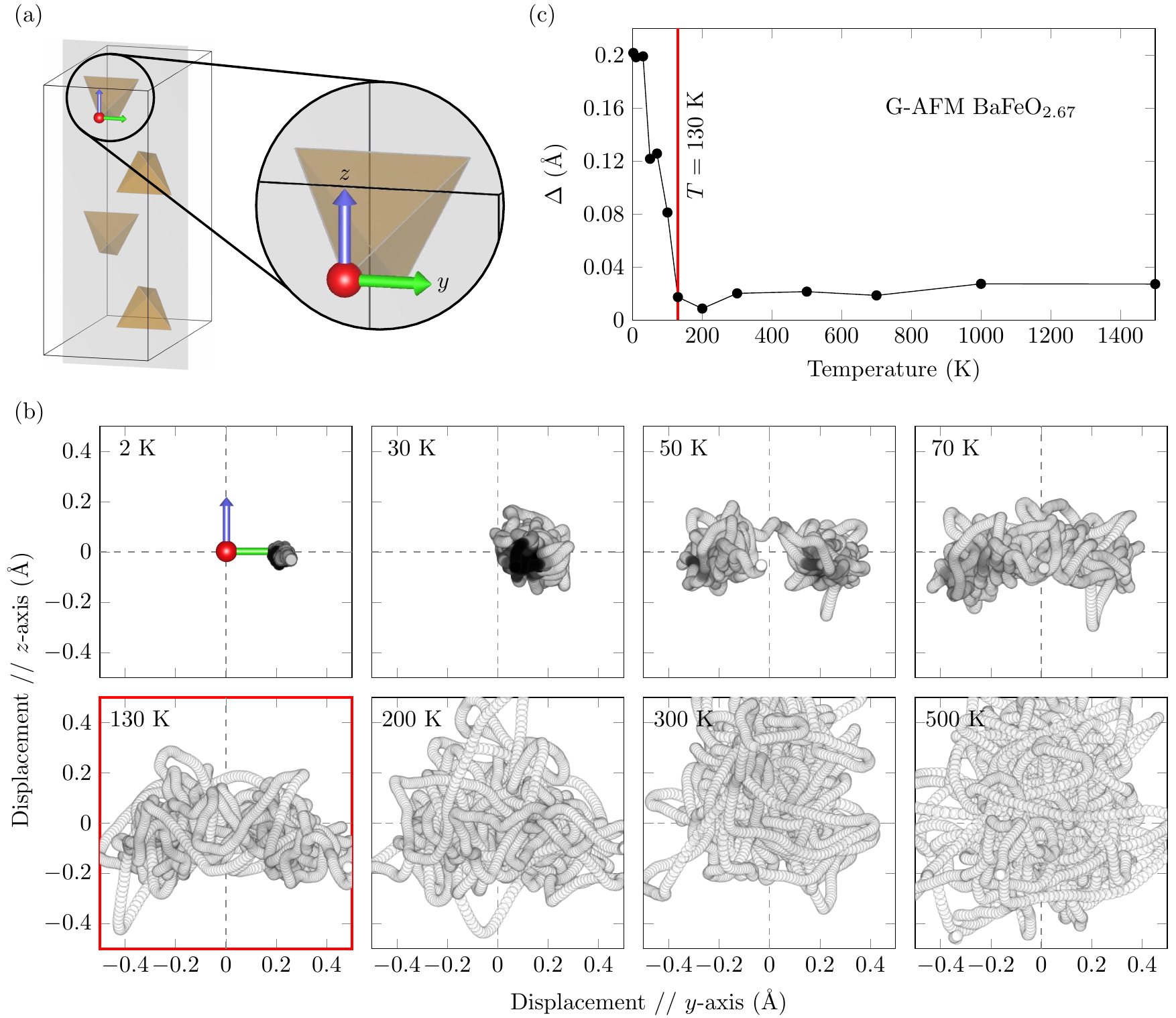}
    \caption{AIMD simulation of the G-AFM vacancy-ordered monoclinic BaFeO$_{2.67}$. (a) Indication of one specific O atom. (b) Trajectories of the O anion at elevated temperatures. (c) The change of the structural descriptor at an elevated temperature. The DFT$+U$ method with $U_{\textrm{eff}} = \SI{3}{\electronvolt}$ was used. }
    \label{fig:monmd}
\end{figure*}

To properly capture the low temperature displacive transformation and the corresponding dynamic stabilization at elevated temperatures, AIMD simulations have been performed from \SI{2}{\kelvin} up to \SI{1500}{\kelvin} for G-AFM BaFeO$_{2.67}$. One of the O atoms involved in the low temperature transformation is focused on for demonstration [Fig.~\ref{fig:monmd}(a)]. Figure~\ref{fig:monmd}(b) shows the trajectories of this O atom at various temperatures. At \SI{2}{\kelvin}, the O atom is trapped in a displaced position in the $y$ direction at a distance of about \SI{0.2}{\angstrom}, which once more substantiates the dynamic instability of the G-AFM monoclinic structure at low temperatures. At \SI{50}{\kelvin} the O atom is able to pass the barrier to the other side within the available simulation time. As the temperature further increases, the O atom can frequently cross the barrier until the trajectory becomes homogeneously spread over a larger region (about \SI{130}{\kelvin}), with no more obvious signs of the displacement (i.e., the average position of the O anion gradually shifts to the ideal position). It is noteworthy that for temperatures \SI{130}{\kelvin} and higher the trajectories become asymmetric along the $z$ axis (which is perpendicular to the double well potential along $y$), with an increased probability to find the O anion at positive $z$ displacements. This finding indicates that the potential energy hypersurface is asymmetric along the $z$ direction.

To quantify the transformation, we utilize again the structure descriptor $\Delta$ extracted from the AIMD simulations. Figure~\ref{fig:monmd}(c) reveals a similar temperature dependence of $\Delta$ as observed for the cubic structure. In particular, the original $P2_{1}/m$ monoclinic structure is stabilized at \SI{130}{\kelvin} due to vibrational entropy. For the ideal monoclinic structure at \SI{0}{\kelvin}, the local magnetic moment for the tetrahedrally coordinated Fe ion is $\SI{3.6}{\mu_{\textrm{B}}}$, and for the other two types of Fe ions are $\SI{3.9}{\mu_{\textrm{B}}}$ and $\SI{4.0}{\mu_{\textrm{B}}}$. The G-AFM state remains stable at all simulated temperatures.

\section{Discussion} \label{sec:dis}

The dynamic instability of the ideal monoclinic BaFeO$_{2.67}$ structure is inherently linked to the ordered vacancies. For the cubic BaFeO$_{3}$  structure, in contrast, all O atoms are symmetrically equivalent and thus contribute equally to the dynamic instability. Due to the formation of the ordered vacancies, four of the twelve Fe coordination polyhedra change from octahedra to tetrahedra, and the site symmetries of the O atoms are modified. For the monoclinic structure, the O atoms near the ordered vacancies mainly contribute to the dynamic instability (Sec.~\ref{sec:mon}). This may be intuited by considering that the O atoms near the ordered vacancies can move more freely than other atoms. The instability may also be related to the Jahn--Teller effect on the Fe$^{4+}$, which is allocated on the tetrahedral site in the previous study~\cite{Wollstadt2021}. 

The ordered vacancies also induce additional anharmonicity for the investigated perovskites. Of course, both the vacancy-free cubic and the vacancy-ordered monoclinic structures are inherently anharmonic along the imaginary modes. As clarified by the double-well potentials [Figs.~\ref{fig:pho3}(e) and \ref{fig:pho2.67}(b)], higher order polynomials with even symmetry are required to stabilize the potential energies. Consistently, the trajectories at temperatures above the transition temperature (about \SI{130}{\kelvin}) show an even-symmetric distribution along the symmetry breaking direction ($x$ in Fig.~\ref{fig:cubmd} and $y$ in Fig.~\ref{fig:monmd}). However, for the vacancy-ordered monoclinic structure an additional anharmonicity in the direction perpendicular to the symmetry-breaking direction is observed ($z$ direction in Fig.~\ref{fig:monmd}). The trajectory distribution along this perpendicular direction is asymmetric, i.e., the O anion prefers on average to be located at positive $z$ displacements. Thus, the creation of ordered vacancies induces an asymmetric local, effective interatomic potential for some O atoms, specifically along the direction perpendicular to the symmetry-breaking direction.


With the DFT$+U$ method, both the cubic and the monoclinic structures are predicted to be stabilized at about \SI{130}{\kelvin}. Based on the comparison of double-well potentials [Fig.~\ref{fig:pho3}(e)], in which deeper potential wells are found for HSE06, a stabilization temperature higher than \SI{130}{\kelvin} is expected for both structures by AIMD with HSE06. 

In contrast, quantum fluctuations, which are not included in AIMD, may decrease the stabilization temperature. In the case of SrTiO$_{3}$, for which displacing along an antiferrodistortive mode was shown to produce an energy decrease of about $\SI{10}{\milli\electronvolt\per{f.u.}}$~\cite{Wahl2008} (depending on the exchange-correlation functional), the transition temperature decreases from \SI{130}{\kelvin} to \SI{110}{\kelvin} due to the quantum fluctuations~\cite{Zhong1996}.

\section{Conclusions} \label{sec:conclusions}

The structures of perovskites with ordered vacancies are much more complex than their vacancy-free cubic counterparts. This challenges an accurate determination of the exact arrangement of atoms. A particularly intricate aspect is the interplay of the ordered vacancies with the dynamic (in)stability of perovskites. In order to gain information on this interplay, we have investigated the dynamic stability of the vacancy-free cubic BaFeO$_{3}$ ($Pm\bar{3}m$) and the vacancy-ordered monoclinic BaFeO$_{2.67}$ ($P2_1/m$) structures with AIMD in this study.

Our results reveal that the ideal monoclinic structure for vacancy-ordered BaFeO$_{2.67}$ is---in contrast to previous expectations~\cite{Wollstadt2021}---dynamically unstable at \SI{0}{\kelvin}. As temperature increases, the ideal monoclinic structure is dynamically stabilized at about \SI{130}{\kelvin}. Interestingly, the ordered vacancies do not significantly alter the critical temperature at which Ba–Fe–O perovskites are dynamically stabilized. The calculated critical temperature is consistent with the dynamic stability requirement in the temperature range at which the high-symmetry phases are experimentally observed (about \SI{300}{\kelvin} to \SI{700}{\kelvin})~\cite{Wollstadt2021}. From a broader perspective, similar results, i.e., dynamical instability at \SI{0}{\kelvin} and stabilization at a relatively low temperature, may also be expected for the vacancy-ordered structures proposed for other perovskite-type phases composed of Ba--Fe--O or even other elements. Such a generalization implies that the dynamic stability is \textit{not} a critical issue for determination of vacancy orderings of perovskites.

Additionally, we have found that strong anharmonicity is induced by the ordered vacancies for the monoclinic structure, along a direction that is perpendicular to the symmetry-breaking direction. This can result in characteristic properties of the material, e.g., a small diffusion barrier of O atoms, which may contribute to the fast ionic diffusion. While the origin of fast ionic diffusion remains an interesting open question~\cite{Wollstadt2021a}, our results provide hints for a possible atomistic mechanism. 

\section*{Acknowledgements} \label{sec:acknowledgements}
This project has received funding from the European Research Council (ERC) under the European Union’s Horizon 2020 research and innovation program (grant agreement No.~865855). The authors also acknowledge support by the state of Baden-Württemberg through bwHPC and the German Research Foundation (DFG) through grant No.~INST~40/467-1~FUGG (JUSTUS cluster) and by the Stuttgart Center for Simulation Science (SimTech).
\appendix*
\section{\texorpdfstring{$U_{\textrm{eff}}$ parameter space}{Ueff parameter space}}\label{sec:pho}

\begin{figure*}[tbp]
    \centering
    \includegraphics[width=17.8cm]{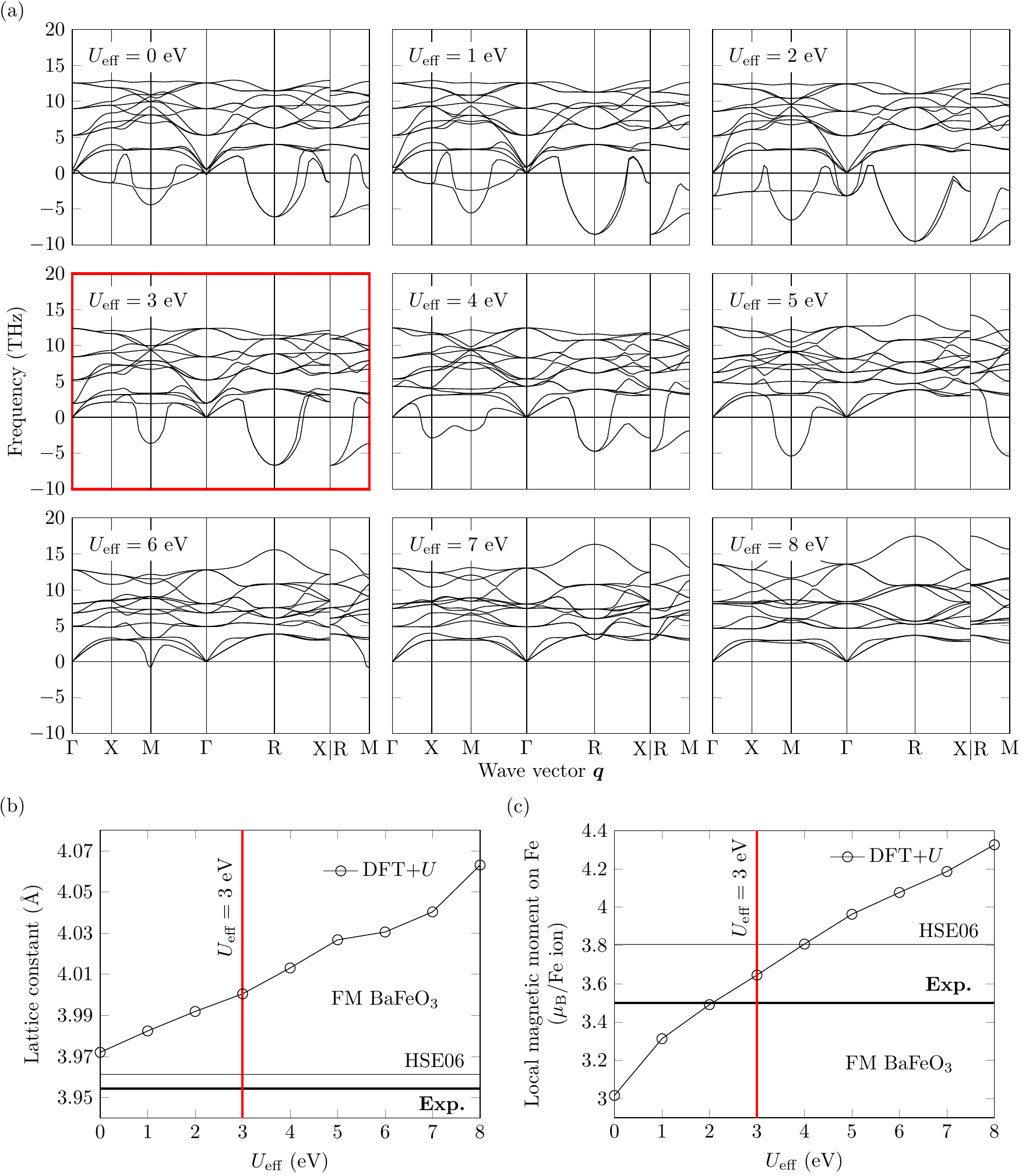}
    \caption{$U_{\textrm{eff}}$ dependence of (a) the phonon dispersion, (b) the lattice constant and (c) the local magnetic moment on Fe for the FM cubic structure BaFeO$_{3}$. Results with $U_{\textrm{eff}} = \SI{3}{\electronvolt}$ are highlighted by a red lines. Phonon dispersion calculated by $U_{\textrm{eff}} = 3$ eV is also shown in Fig.~\ref{fig:pho3}(b). Commensurate $\boldsymbol{q}$-points are indicated by tick labels. Imaginary modes are shown by negative frequencies. Experimental data are obtained from Ref.~\cite{Hayashi2011}. }
    \label{fig:u}
\end{figure*}

Results calculated by the DFT$+U$ method are sensitive to the input parameter $U_{\textrm{eff}}$, as shown by Table~\ref{tab:pre}. We have explored the DFT$+U$ parameter space by testing a range of $U_{\textrm{eff}}$ values from \SI{0}{\electronvolt} to \SI{8}{\electronvolt} for the FM cubic BaFeO$_{3}$. Figure~\ref{fig:u}(a) shows the $U_{\textrm{eff}}$ dependence of the phonon dispersion. With a relatively small $U_{\textrm{eff}}$ value ($\leq \SI{2}{\electronvolt}$), imaginary modes (which indicate dynamic instability at \SI{0}{\kelvin}) are found at all commensurate $\boldsymbol{q}$-points included here, i.e., $\Gamma$, M, R, X. Less imaginary modes are present at larger $U_{\textrm{eff}}$. Specifically, imaginary modes at the M point disappear when $U_{\textrm{eff}} \approx \SI{6}{\electronvolt}$ while at the R point they disappear when $U_{\textrm{eff}} \approx \SI{5}{\electronvolt}$. No imaginary mode is found for phonon calculations with a $U_{\textrm{eff}}$ value larger than \SI{7}{\electronvolt}. The result for $U_{\textrm{eff}} = \SI{5}{\electronvolt}$ is found to be different from previous studies~\cite{Cherair2017, Zhang2018a} (shown in Table~\ref{tab:pre}), which may be due to the usage of different functionals or codes. 

The change of the lattice constant is shown in Fig.~\ref{fig:u}(b). With $U_{\textrm{eff}} = \SI{0}{\electronvolt}$ (GGA functional), the calculated value is higher than the experimental value~\cite{Hayashi2011} (about 0.5\%). The overestimation of the lattice constant calculated by the GGA functional is also shown for the SrTiO$_{3}$ and the BaTiO$_{3}$ cubic perovskites~\cite{Wahl2008}. The DFT$+U$ method increases the amplitude of the overestimation. With $U_{\textrm{eff}} = \SI{3}{\electronvolt}$, the lattice constant is about 1\% larger than the experimental value. The change of the local magnetic moment on Fe is shown in Fig.~\ref{fig:u}(c). The GGA functional underestimates the local magnetic moment by about 15\%, which is corrected by the DFT$+U$ method. With $U_{\textrm{eff}} = \SI{3}{\electronvolt}$, the calculated local magnetic moment is between the values given by experiments~\cite{Hayashi2011} and HSE06 (calculated in this study).

\newpage
\bibliography{ref}

\end{document}